# Field Localization and Enhancement of Phase Locked Second and Third Harmonic Generation in Absorbing Semiconductor Cavities

V. Roppo[1,5], C. Cojocaru[1], F. Raineri[2,3], G. D'Aguanno[5] J. Trull[1], Y. Halioua[2,4], R. Raj[2], I. Sagnes[2], R. Vilaseca[1], M. Scalora[5]

[1]*Universitat Politècnica de Catalunya, Colom 11, E-08222 Terrassa, Spain*

[2]*Laboratoire de Photonique et de Nanostructures (CNRS UPR20), Route de Nozay, 91460 Marcoussis, France*

[3]*Université Paris-Diderot, 75205 Paris Cedex 13, France*

[4]*Photonics Research Group, Department of Information Technology, Ghent University B-9000 Ghent, Belgium*

[5]*Charles M. Bowden Research Facility, US Army RDECOM, Redstone Arsenal AL 35803, USA*

**Abstract**

We predict and experimentally observe the enhancement by three orders of magnitude of phase mismatched second and third harmonic generation in a GaAs cavity at 650nm and 433nm, respectively, well above the absorption edge. Phase locking between the pump and the harmonics changes the effective dispersion of the medium and inhibits absorption. Despite hostile conditions the harmonics become localized inside the cavity leading to relatively large conversion efficiencies. Field localization plays a pivotal role and ushers in a new class of semiconductor-based devices in the visible and UV ranges.

Since it was discovered by Franken in the 1960's, second harmonic (SH) generation has been one of the most studied phenomena in nonlinear optics [1]. To date most efforts have been directed at improving the efficiency of the process by developing new materials with high effective nonlinear coefficients, accompanied by phase and group velocity matching [2-10]. Consequently, most studies have been concerned with maximizing conversion efficiencies, generally achievable at or very near phase matching (PM) conditions, ensuring maximum energy transfer from the fundamental beam to the harmonics. A special effort was focused toward engineering new artificial materials capable of compensating material dispersion, for example, using quasi phase matching technique [11, 12] or structured materials [13]. Outside of PM conditions and low conversion efficiency [3], the only relevant processes that have been investigated are cascaded parametric processes that can produce phase-modulation of the fundamental beam [14], pulse breaking [15] or nonlinear diffraction [16]. This has caused other possible working conditions to remain largely unexplored. A relevant feature is that in all these previous studies the nonlinear material was assumed to be transparent for both fundamental and harmonics beams, since conventional wisdom holds that an absorptive material will reabsorb any generated harmonic signal.

More recently, an effort was initiated to systematically study the behavior of SH and third harmonic (TH) fields in transparent and opaque materials under conditions of phase mismatch [17-19]. Briefly, when a pump pulse crosses an interface between a linear and a nonlinear medium there are always three generated SH (and/or TH) components. One component is generated backward into the incident medium, due the presence of the interface, and the remaining components are forward-moving. These components may be understood on the basis of the mathematical solution of the homogeneous and inhomogeneous wave equations at the SH frequency [4]. The two forward-propagating components interfere in the vicinity of the entry surface and give rise to Maker fringes [2, 20] and to energy exchange between the fundamental and SH and/or TH beams. It turns out that while the homogeneous component travels with the group velocity given by material dispersion, the inhomogeneous component is captured by the pump pulse and experiences the same effective dispersion of the pump [21]. That is, the homogenous component has wave-vector $k_{2\omega}^{HOM} = n(2\omega)k(2\omega)$ and exchanges energy with the pump until the inevitable walk-off. The inhomogeneous, phase-locked (PL) component has a wave-vector $k_{2\omega}^{PL} = 2n(\omega)k(\omega)$, twice the pump wave-vector, and travels locked to the pump pulse.

The consequences of phase locking can guide us towards new scenarios by allowing working conditions hitherto assumed inaccessible for absorbing materials, semiconductors in particular. In ref.[19] it was shown that in the opaque region inhomogeneous SH and TH components are not absorbed. This behaviour may be understood within the frame-work of the phase locking mechanism. The real part of the effective index of refraction that the harmonics experience is equal to that of the pump [17]. This result follows from a full spectral decomposition of the wave packets in $k$ and $\omega$ spaces, as outlined in ref.[21]. Then, a Kramers-Kronig reconstruction of the effective index leads to a complex effective index at the harmonic wavelengths that is identical to the pump's complex index of refraction. It naturally follows that the suppression of absorption at the harmonic wavelengths will occur if the pump is tuned to a region of transparency. The results reported in refs.[19, 21] are concrete evidence that phase and group velocity locking leads to the inhibition of linear absorption. The next step is to learn how to manage and exploit this ubiquitous phenomenon, possibly in spectral regions previously thought to be inaccessible.

In this Letter we highlight the surprising behavior of SH and TH phase-locked components with frequencies above the absorption edge by showing that, when the material is placed inside a cavity resonant only at the fundamental frequency, the PL mechanism not only inhibits absorption but also fosters the enhancement of harmonic generation by several orders of magnitude compared to the no-cavity case. Our interpretation is that this enhancement arises because of two complementary factors. First, phase locking with the resonant pump pulls the harmonics into effective resonance too, leading to field enhancement and increased energy exchange between the fields. In this regard, we note that the rate at which energy is transferred from a nonlinear source to a field at a harmonic frequency (for instance at frequency $2\omega$) is proportional to $\mathbf{J}_{2\omega} \bullet \mathbf{E}_{2\omega}$ at each point inside the material, where $\mathbf{J}_{2\omega} = -\partial \mathbf{P}_{2\omega}/\partial t$ is the current density and $\mathbf{P}_{2\omega}$ is the harmonic polarization induced by the fundamental field. Since currents, polarization, and fields are local variables, the conversion efficiency depends on the strengths of both the FF and SH (or TH) fields at each point inside the cavity. Second, if the cavity filled with the nonlinear medium is short (only a few FF wavelengths thick), the fundamental beam inside the medium "visits" the front and rear interfaces many times during the duration of a light pulse. As shown in [17-19] and again as pointed out above, in the opaque region it is just near the interfaces that energy can flow from the pump to the harmonic fields (far from the interfaces, once the homogenous component is absorbed, the total energy of the phase-locked component clamps and remains constant [17]).

This increase in the conversion efficiency within a cavity can be illustrated with a simple example, consisting of a thin layer, or free-standing etalon, of a semiconductor material, for instance GaAs, with an optical thickness of only two times the FF wavelength. This is a cavity with a relatively small Q factor. Let us assume a fundamental beam is tuned at 1300 nm and generates SH (650 nm) and TH (433 nm) signals. The dispersion curves for bulk GaAs (Fig.1(a)) show that the material is transparent above 900nm, and opaque below900nm. The complex refractive index of GaAs is $n(1300\text{nm})=3.41$, $n(650\text{nm})=3.83+i0.18$ and $n(433\text{nm})=5.10+i1.35$, as reported by Palik [22]. The GaAs Fabry-Perot etalon length (760 nm) is such that the sample is resonant at the pump wavelength, transmits ~3% of the incident light at 650nm, and is completely opaque at 433nm. These conditions mean that the generated SH field is dominated by the PL component, but contains some residual homogeneous signal as well and is mostly in phase with the FF field. In contrast, the TH field will contain only the phase locked component, as revealed by a k-space spectral analysis of the field eigenmodes. The thin blue curve in Fig. 1(b) shows the standing-wave profile of the fundamental magnetic field intensity inside the etalon. The thick red curve shows the corresponding generated SH magnetic field intensity profile. Ostensibly, the SH field pattern is in phase with that of the FF. This situation should be contrasted with the SH field generated when absorption is artificially turned off (black dashed curve). In the latter case the SH field has an odd number of peaks and does not resonate in phase with the FF due to interference between homogeneous and PL components. Although not pictured, the TH undergoes similar dynamics. Thus the PL mechanism causes the inhomogeneous components to resonate inside the cavity along with the pump, *regardless of material dispersion at the harmonic wavelengths* in what one might characterize as a double action of fundamental field and *anomalous harmonic field localization*. Of course, the phenomenon occurs with any nonlinear absorbing material, including negative index materials and semiconductors in the metallic range [23].

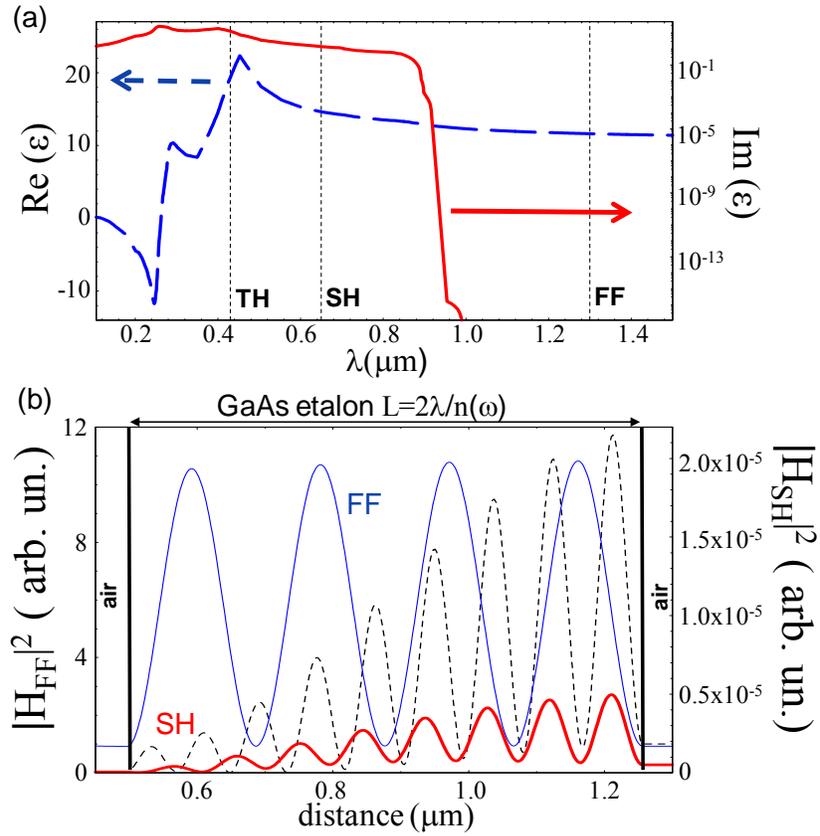

Figure 1: (Color online) (a) Dispersion for bulk GaAs. (b) Numerical calculation of the localization of the H fields for the fundamental (FF) (thin, blue solid curve) and SH (thick, red solid curve) in an etalon two wavelengths thick. We also plot the SH field for the case where the absorption of the material is turned off (thin dashed curve). The incident field is Gaussian in shape and 80fs in duration.

The conversion efficiency in the case of a simple etalon increases by several orders of magnitude with respect to the case of a bulk medium, in spite of the fact that the Q factor of the cavity is relatively small. Next, we show that slight improvements to the cavity, for example, by adding a mirror at the back interface, conversion efficiencies increase even more with respect to a bulk medium. We consider a GaAs layer with a gold mirror in the back. The calculations to optimize the GaAs cavity with the gold mirror for SH and TH emission were performed using a plane wave approach based on the Green function for multilayered structures developed in ref. [24]. In Fig.2 we show the results of this theoretical study for an input intensity of ~5GW/cm$^2$ and an incident wavelength 1300nm, and compare to the conversion efficiency of the Fabry-Perot etalon we discussed earlier. We assume $\chi^{(2)}$~14pm/V and $\chi^{(3)}$~1.7x10$^{-19}$m$^2$/V$^2$ in all cases. The reason for these choices will become clear later. Fig.2(a) shows the SH conversion efficiency (dashed blue curve) as a function of cavity length, L. The peaks in the SH efficiency curve appear at the cavity lengths that make the FF field resonant. An absolute maximum is found for L~645 nm, with conversion efficiency of ~10$^{-6}$. The peaks are slowly modulated by the residual presence of homogenous SH components. The TH conversion efficiency curve in Fig.2b shows no such modulation, an indication that the TH signal is completely phase-locked. In comparison, in Figs.2 the thin (green) and thick (red) curves represents transmitted and reflected efficiencies, respectively, for SH (Fig.2(a)) and TH (Fig.2(b)) for a free-standing GaAs etalon of similar thickness. It is evident from Fig.2 that the configuration with the back-mirror yields nonlinear conversion efficiencies one order of magnitude higher compared to a free-standing GaAs etalon, due to enhanced field localization. At the same time, and even more remarkably, conversion efficiencies are nearly *four* orders of magnitude larger compared to the case of a GaAs bulk medium, which are of order $10^{-10}$ for both SH and TH for an input intensity of a few GW/cm$^2$ and an incidence angle of ~10°. This reference value was taken from our experiments on a SH signal generated from a GaAs substrate, and is of the same level as background noise.

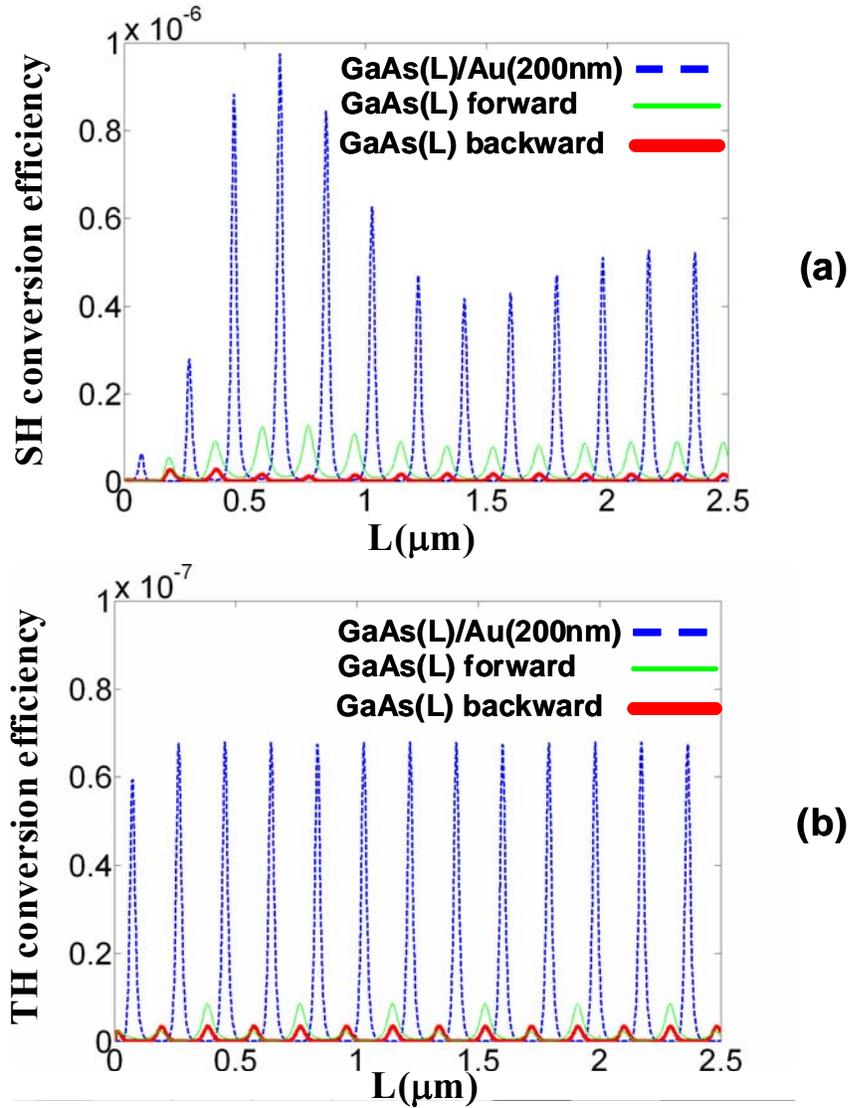

Figure 2: SH (a) and TH (b) conversion efficiencies vs GaAs cavity length with (dashed blue curves) and without (simple etalon, green and red curves) a back mirror. Both forward and backward generation results from the etalon.

We now present experimental results for a structure that we fabricated and tested in the laboratory. It corresponds to the same configuration consisting of a GaAs layer having a back gold mirror. The sample was fabricated using MOCVD to grow a 645nm GaAs layer above a AlAs etch-stop layer, on top of a GaAs substrate. A gold mirror approximately 200nm thick was deposited onto the GaAs layer, aided by a few nanometers of Ti buffer layer to insure good adhesion between the metal and the semiconductor (schematization in Fig. 3(a)). The structure was then glued upside down onto a silicon substrate using benzocyclobutene polymer (Fig. 3(a)). Finally, the GaAs substrate and AlAs layer were removed using mechanical grinding followed by chemical etching. The linear reflectance, Fig.3(b), shows that the structure displays a resonance at the FF, 1300nm. However, the stack is not resonant at the SH and TH wavelengths. We note that the introduction of the thin Ti layer spoils cavity conditions for the pump, absorbing part of it and decreasing the total reflectance of the structure at the FF frequency from approximately 80% (without the Ti layer) down to the ~13% (with the Ti layer) shown in Fig. 3(b).

In our experiment we used a typical reflection measurement set-up, shown schematically in Fig.3(a). The source consists of (80-120)fs fundamental pulses from an OPA laser working at 1KHz repetition rate, with tunable wavelength between 1200 and 1600nm. The beam has a power of 200MW and was focused on the sample down to a ~0.5 millimeter spot size, with corresponding peak intensity of ~5GW/cm$^2$. The reflected signal (Fig. 3(a) was collected and analyzed with a spectrometer connected to a cooled Si CCD camera. The experiment consisted of scanning the sample's resonance at 1300nm with the fundamental pulse laser, and retrieving SH and TH signals. Nine different measurements were performed from 1260nm up to 1420nm in 20nm wavelength steps. As references, the SH and TH signals generated from a simple Au mirror and a bulk GaAs sample, as well as the background illumination, were recorded with the same set-up and subtracted from the harmonic signals recorded with our sample. These references show clearly that surface SH and TH signals generated by the bulk GaAs and gold samples are negligible with respect to the harmonics generated by the cavity. The SH and TH measured for each step of the fundamental tuning are shown in Fig.3(c) and 3(d), respectively. The vertical axis shows the conversion efficiency of each process. These results show that the maximum SH efficiency occurs at ~650nm and for the TH at ~435nm: this is remarkable proof that the harmonics display resonant behavior. The dashed curve represents the envelope of the fields obtained numerically in the continuous wave (CW) regime for the cavity in our experiment, and the agreement is very good when $\chi^{(2)} \sim 14$pm/V and $\chi^{(3)} \sim 1.7 \times 10^{-19}$m$^2$/V$^2$. Due to field enhancement and overlap this time we recorded a SH conversion efficiency of order $1.5 \times 10^{-7}$. The presence of the Ti layer, introduced as practical solution to a mechanical gold adhesion problem, unfortunately also results in conversion efficiencies that are approximately six times smaller than what is actually possible were the Ti layer not present (Fig 2a). However, even under these conditions conversion efficiencies are at least three orders of magnitude larger than the SH signal generated by bulk GaAs under the same conditions. Even more fascinating is the TH situation, where we recorded efficiencies of order of $1.4 \times 10^{-8}$ under conditions of even higher absorption and "wrong" cavity length. This is testament to the robustness of the phase-locking mechanism. Additionally to the simple illustrative examples discussed above, preliminary calculations with higher-Q cavities show that the same GaAs etalon sandwiched between distributed feedback mirrors (i.e. a photonic band gap structures with defect states) yields conversion efficiencies that rival and may even surpass phase-matched SHG conversion efficiencies, once again thanks to a combination of significant local field enhancement and localization between the fundamental and harmonic fields.

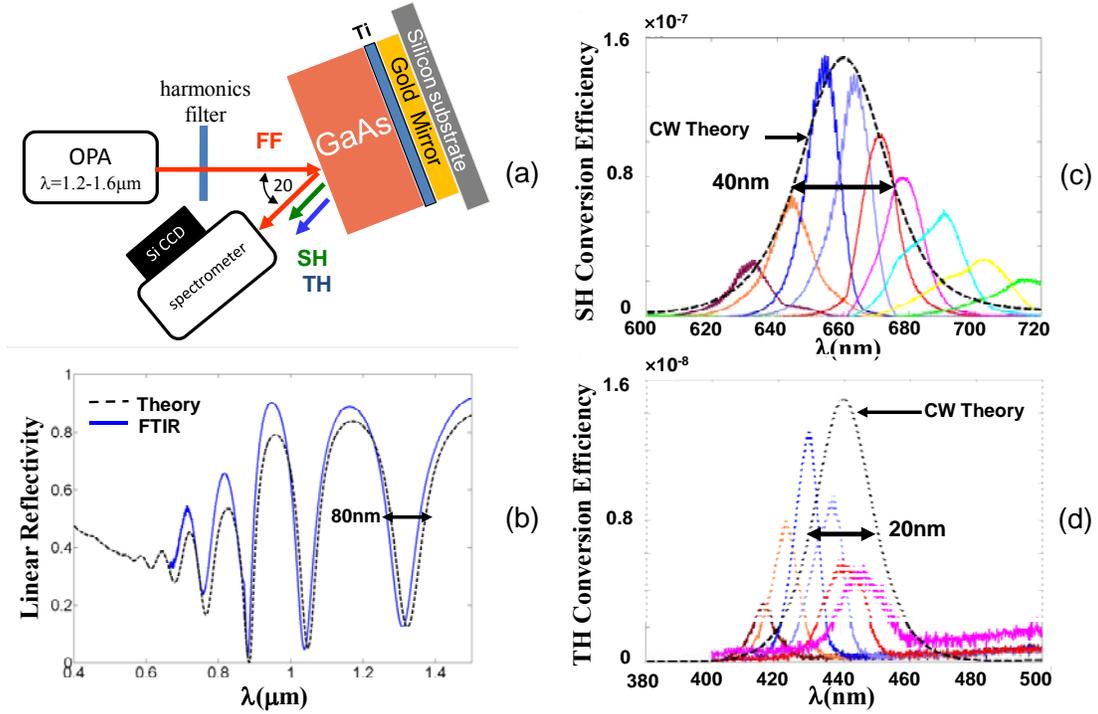

Figure 3:(Color online) (a) Schematic drawing of the experimental set-up and the sample. (b) Linear reflectivity of the sample taken by Fourier Transform Infrared Spectroscopy (FTIR) and comparison with theory. Experimental results for SH (c) and TH (d) signals. The dashed black curves represent the CW theory. The different curves represent the harmonic signals for different pump tunings (respectively from left to right 1260nm, 1280nm, 1300nm, 1320nm, 1340nm, 1360nm, 1380nm, 1400nm, 1420nm). We note that the bandwidths of the signals are in excellent agreement with the numerical predictions.

As a rough estimate of what one might expect in terms of conversion efficiency Fermi's golden rule provides a good practical procedure to follow. According to this rule, the spontaneous emission rate is: $\gamma = \frac{2\pi}{\hbar} \rho(\omega) |<f|\mathbf{\mu} \cdot \mathbf{E}|i>|^2$, where $|f>$ and $|i>$ are final and initial states, respectively, $\rho(\omega)$ is the density of states, $\mathbf{\mu}$ is the dipole moment, and $\mathbf{E}$ is the local electric field. Both $\rho(\omega)$ and $|\mathbf{E}|^2$ are proportional to the cavity Q. As a result nonlinear conversion rates are proportional to $Q^2$. One should keep in mind that these estimates are just that, and that geometrical factors like field localization, dipole position and distribution inside the cavity intervene to alter these estimates through shape factors.

In summary, we have shown here, theoretically and experimentally, that a low-Q GaAs cavity resonant only with the pump can yield SH and TH conversion efficiencies at least three orders of magnitude larger compared to bulk GaAs, and can be dramatically improved if the cavity quality factor is increased. Transparency and cavity resonance to the FF combined with cavity effects are the key factors regardless of absorption and resonance conditions at the SH and TH frequencies, since the PL SH and TH components experience the material dispersion/absorption characteristics of the pump wavelength. These results have general validity, and apply well to semiconductors and dielectric materials alike, as well as negative index materials, and open the door to the discovery and examination of new optical phenomena in wavelength ranges that are far below the absorption edge, well into the metallic region of semiconductors [23]. These results are bound to find straightforward applications, for example, in new low cost, easily tunable and easily fabricated UV and X-UV sources.

**Acknowledgments:** We thank the US Army European Research office (project W911NF) and the Spanish government (FIS2008-06024-C03-02) for partial financial support. G.D. thanks the National Research Council for financial support. We also thank Nadia Mattiucci and Mark J. Bloemer for helpful discussions and suggestions.